# Regularized Consensus PCA

**Michel Tenenhaus (1), Arthur Tenenhaus (2), Patrick J. F. Groenen (3)**

(1) HEC Paris, Jouy-en-Josas, France, tenenhaus@hec.fr

(2) CentraleSupélec-L2S, UMR CNRS 8506, Gif-sur-Yvette, France and Bioinformatics-Biostatistics Platform IHU-A-ICM, Paris, France, arthur.tenenhaus@centralesupelec.fr

(3) Econometric Institute, Erasmus School of Economics, Erasmus University, Rotterdam, groenen@ese.eur.nl

## Abstract

A new framework for many multiblock component methods (including consensus and hierarchical PCA) is proposed. It is based on the consensus PCA model: a scheme connecting each block of variables to a superblock obtained by concatenation of all blocks. Regularized consensus PCA is obtained by applying regularized generalized canonical correlation analysis to this scheme for the function $g(x) = x^m$ where $m \geq 1$. A gradient algorithm is proposed. At convergence, a solution of the stationary equation related to the optimization problem is obtained. For $m$ = 1, 2 or 4 and shrinkage constants equal to 0 or 1, many multiblock component methods are recovered.

**Keywords**: Consensus PCA, hierarchical PCA, multiblock component methods, regularized generalized canonical correlation analysis.

## 1. Introduction

In this paper, we consider several data matrices $\mathbf{X}_1, ..., \mathbf{X}_b, ..., \mathbf{X}_B$. Each $n \times J_b$ data matrix $\mathbf{X}_b$ is called a block and represents a set of $J_b$ variables observed on $n$ individuals. The number and the nature of the variables usually differ from one block to another but the individuals must be the same across blocks. We assume that all variables are centered. Generalized canonical correlation analysis (Carroll, 1968a), consensus PCA (Westerhuis, Kourti, and MacGregor, 1998) and hierarchical PCA (Smilde, Westerhuis, and de Jong, 2003) can be considered as methods extending PCA of variables to PCA of blocks of variables. These methods belong to the family of multiblock component methods. These three methods are special cases of a very general optimization problem. The objective is to find block components $\mathbf{y}_b = \mathbf{X}_b\mathbf{w}_b$, $b = 1, ..., B$ and an auxiliary variable $\mathbf{y}_{B+1}$ solution of the following optimization problem:

(1)
$$\underset{\mathbf{w}_1,...,\mathbf{w}_B,\mathbf{y}_{B+1}}{\text{Maximize}} \sum_{b=1}^{B} \left(\text{cov}(\mathbf{X}_b\mathbf{w}_b, \mathbf{y}_{B+1})\right)^m$$

$$\text{s.t. } \mathbf{w}_b^t \mathbf{M}_b \mathbf{w}_b = 1, \ b = 1,...,B, \text{ and } \text{var}(\mathbf{y}_{B+1}) = 1$$

where $m$ is any value $\geq 1$ and $\mathbf{M}_b = \tau_b \mathbf{I} + (1-\tau_b)\frac{1}{n}\mathbf{X}_b^t\mathbf{X}_b$ with a shrinkage constant $\tau_b$ between 0 and 1. The resulting auxiliary variable $\mathbf{y}_{B+1}$ plays the role of a principal component of the superblock taking into account the block structure. The quantity $c_b = \left(\text{cov}(\mathbf{X}_b\mathbf{w}_b, \mathbf{y}_{B+1})\right)^m / \sum_{b=1}^{m} \left(\text{cov}(\mathbf{X}_b\mathbf{w}_b, \mathbf{y}_{B+1})\right)^m$ measures the contribution of block $b$ to the construction of the auxiliary variable $\mathbf{y}_{B+1}$. For $m = 2$ and all $\tau_b = 0$, (1) corresponds to Carroll's generalized canonical correlation analysis. For $m = 2$ and all $\tau_b = 1$, consensus PCA (version of Westerhuis, Kourti, and MacGregor, 1998) is recovered; in this case, the auxiliary variable is in fact the first principal component of the superblock. For $m = 4$ and all $\tau_b = 1$, hierarchical PCA (version of Smilde, Westerhuis, and de Jong, 2003) is obtained.

In this paper, we show that the three multiblock component methods mentioned above and many others can also be included in a general framework that we call "regularized consensus PCA". It is based on the "consensus PCA model" defined as a scheme connecting each block $\mathbf{X}_1, ..., \mathbf{X}_b, ..., \mathbf{X}_B$ to the superblock $\mathbf{X}_{B+1} = [\mathbf{X}_1, ..., \mathbf{X}_b, ..., \mathbf{X}_B]$. In Wold (1982) and in Lohmöller (1989) this model is discussed under the name "hierarchical model with repeated manifest variables", but in this paper we prefer to use the first name. In (1), the auxiliary variable is replaced by a superblock component $\mathbf{y}_{B+1} = \mathbf{X}_{B+1}\mathbf{w}_{B+1}$ and the following optimization problem is considered:

(2)
$$\underset{\mathbf{w}_1,...,\mathbf{w}_{B+1}}{\text{Maximize}} \sum_{b=1}^{B} \left(\text{cov}(\mathbf{X}_b\mathbf{w}_b, \mathbf{X}_{B+1}\mathbf{w}_{B+1})\right)^m$$
$$\text{s.t. } \mathbf{w}_b^t \mathbf{M}_b \mathbf{w}_b = 1, \ b = 1, ..., B+1$$

where $m \geq 1$, $\mathbf{M}_b = \tau_b \mathbf{I} + (1 - \tau_b)\frac{1}{n}\mathbf{X}_b^t \mathbf{X}_b$ and $0 \leq \tau_b \leq 1$. Optimization problem (2) is a special case of RGCCA-M, a new regularized generalized canonical correlation analysis method proposed by Groenen and Tenenhaus (2015). In this paper, (2) will be studied directly without reference to RGCCA-M because many simplifications occur and new mathematical properties are obtained when considering this specific problem. In Section 2, we present a very simple iterative algorithm (called R-CPCA) related to (2). It is proved that this algorithm is monotonically convergent and yields a solution of the stationary equation related to (2). Furthermore, under a mild condition, the sequences of block and superblock weight-vectors are converging to this solution. However, it is not guaranteed that the globally optimal solution of (2) is reached, although this seems to be the case in practical applications.

Many well-known multiblock component methods are obtained as special cases of (2) by fixing $m$ and the shrinkage constants $\tau_1, ..., \tau_B, \tau_{B+1}$. In Section 3, we study various special cases of the R-CPCA algorithm for $m = 1, 2$ or $4$. When all shrinkage constants $\tau_b$ are equal to 0, several correlation-based methods are recovered and when all these $\tau_b$ are equal to 1, covariance-based methods are recovered. For $\tau_1 = ... = \tau_B = 1$ and $\tau_{B+1} = 0$, or the opposite, generalizations of redundancy analysis are obtained. Using shrinkage constants between 0 and 1 insures a continuum between covariance-based and correlation-based methods.

Higher order block and superblock components can be obtained by adding orthogonality constraints to (2). Various deflation strategies are discussed in Section 4. Regularized consensus PCA allows the construction of orthogonal components in each block and in the superblock. In the usual consensus PCA and related methods, this orthogonality property of both the block and superblock components is not achievable because these methods use an auxiliary variable and not a superblock.

In Section 5, we discuss the connections between many multiblock component methods, regularized generalized canonical correlation analysis (Tenenhaus and Tenenhaus, 2011) and PLS path modeling (Wold (1982, 1985), Lohmöller (1989), Tenenhaus, Esposito Vinzi, Chatelin and Lauro (2005)).

## 2. Regularized consensus PCA

Consensus PCA was first introduced by Wold, Hellberg, Lundstedt, Sjöström and Wold (1987) as a method extending PCA of variables to PCA of blocks of variables. Westerhuis, Kourti, and MacGregor (1998) noticed some convergence problem in the consensus PCA algorithm and proposed a slightly modified version which solved this problem. Then, Hanafi, Kohler and Qannari (2011) showed that this modified algorithm was actually maximizing a covariance-based criterion. This modified consensus PCA algorithm and many other multiblock component methods are recovered with regularized consensus PCA.

### 2.1 Definition of regularized consensus PCA

Regularized consensus PCA of a set of data matrices $\mathbf{X}_1, ..., \mathbf{X}_b, ..., \mathbf{X}_B$ observed on the same individuals is defined by optimization problem (2). The choice $\mathbf{M}_b = \mathbf{I}$ corresponding to $\tau_b = 1$ is called Mode A while the choice $\mathbf{M}_b = (1/n)\mathbf{X}_b^t\mathbf{X}_b$ corresponding to $\tau_b = 0$ is called Mode B. Choosing a shrinkage constant $\tau_b$ between 0 and 1 yields a shrinkage estimate $\mathbf{M}_b = \tau_b \mathbf{I} + (1-\tau_b)(1/n)\mathbf{X}_b^t\mathbf{X}_b$ of the true covariance matrix $\mathbf{\Sigma}_b$ related to block $b$ (Ledoit and Wolf, 2004) and insures a continuum between Modes A and B. Note that the inverse of $\mathbf{M}_b$ is used in Algorithm 1 presented below to solve (2). Therefore, when Mode B is selected in (2) for a block or superblock data matrix $\mathbf{X}_b$, we assume that this matrix is full rank. Note that when the superblock is full rank and Mode B selected for this superblock, problems (1) and (2) give the same results. This point will be discussed below.

How to manage a situation where the user wants to select Mode B for not full rank block or superblock? We first consider a not full rank superblock. Mode B cannot be selected for this superblock in (2), but (1) can still be used instead. Regularized consensus PCA can also be extended to a situation where Mode B is selected for a not full rank block $\mathbf{X}_b$ (but with $\text{rank}(\mathbf{X}_b) < n$, otherwise regularization is mandatory) by replacing in the criterion used in (2) $\mathbf{X}_b\mathbf{w}_b$ by $\mathbf{y}_b$ and by imposing the constraints $\mathbf{y}_b = \mathbf{X}_b\mathbf{w}_b$ and $\text{var}(\mathbf{y}_b) = 1$. This situation will be discussed at the end of Section 2.

## 2.2 Regularized consensus PCA of transformed data

Optimization problem (2) can be simplified. Let $\mathbf{P}_b = \mathbf{X}_b \mathbf{M}_b^{-1/2}$, $\mathbf{Q}_b = \mathbf{P}_b^t \mathbf{P}_{B+1}$, $\mathbf{v}_b = \mathbf{M}_b^{1/2} \mathbf{w}_b$ and $\mathbf{y}_b = \mathbf{X}_b \mathbf{w}_b = \mathbf{P}_b \mathbf{v}_b$. Expressing (2) in terms of $\mathbf{P}_b$ and $\mathbf{v}_b$ yields

$$\text{(3)} \quad \underset{\mathbf{v}_1,\ldots,\mathbf{v}_{B+1}}{\text{Maximize}} \sum_{b=1}^{B} \left( \text{cov}(\mathbf{P}_b \mathbf{v}_b, \mathbf{P}_{B+1} \mathbf{v}_{B+1}) \right)^m$$

$$\text{s.t. } \mathbf{v}_b^t \mathbf{v}_b = 1, \ b = 1,\ldots, B+1.$$

Maximizing each term of the criterion in (3) for a fixed $\mathbf{v}_{B+1}$ gives

$$\text{(4)} \quad \mathbf{v}_b^* = \mathbf{P}_b^t \mathbf{P}_{B+1} \mathbf{v}_{B+1} / \left\| \mathbf{P}_b^t \mathbf{P}_{B+1} \mathbf{v}_{B+1} \right\| = \mathbf{Q}_b \mathbf{v}_{B+1} / \left\| \mathbf{Q}_b \mathbf{v}_{B+1} \right\|$$

This computation assumes $\mathbf{Q}_b \mathbf{v}_{B+1} \neq \mathbf{0}$. In fact, $\mathbf{Q}_b \mathbf{v}_{B+1} = \mathbf{P}_b^t \mathbf{P}_{B+1} \mathbf{v}_{B+1} = \mathbf{0}$ implies that $\text{cov}(\mathbf{X}_b \mathbf{w}_b, \mathbf{y}_{B+1}) = 0$ for any $\mathbf{w}_b$. This means that the block $\mathbf{X}_b$ does not contribute to the superblock component $\mathbf{y}_{B+1}$ and could be removed from the analysis. Therefore, we suppose in the following that all $\left\| \mathbf{Q}_b \mathbf{v}_{B+1} \right\|$ appearing in this paper are strictly positive. For the optimal $\mathbf{v}_b^*$ in (4), we obtain

$$\text{(5)} \quad \text{cov}(\mathbf{P}_b \mathbf{v}_b^*, \mathbf{P}_{B+1} \mathbf{v}_{B+1}) = \frac{1}{n} \left\| \mathbf{Q}_b \mathbf{v}_{B+1} \right\|.$$

Using (5), optimization problem (3) becomes

$$\text{(6)} \quad \text{Maximize } \frac{1}{n^m} \sum_{b=1}^{B} \left\| \mathbf{Q}_b \mathbf{v}_{B+1} \right\|^m \text{ s.t. } \left\| \mathbf{v}_{B+1} \right\| = 1.$$

Note that the objective function $\Psi(\mathbf{v}) = \sum_{b=1}^{B} \left\| \mathbf{Q}_b \mathbf{v} \right\|^m$ is convex and continuously differentiable for $m \geq 1$. Convexity comes from the fact that the Euclidean norm is convex and that a monotonically increasing convex function of a convex function is convex, too. Therefore, we may use all the results on the maximization of a convex function presented in Appendix 1.

We first compute the gradient $\nabla(\mathbf{v})$ of $\Psi(\mathbf{v})$:

$$\text{(7)} \quad \nabla(\mathbf{v}) = \sum_{b=1}^{B} \frac{\partial \left\| \mathbf{Q}_b \mathbf{v} \right\|^m}{\partial \mathbf{v}} = \frac{m}{2} \sum_{b=1}^{B} \left( \left\| \mathbf{Q}_b \mathbf{v} \right\|^2 \right)^{\frac{m}{2}-1} \frac{\partial \left\| \mathbf{Q}_b \mathbf{v} \right\|^2}{\partial \mathbf{v}} = m \sum_{b=1}^{B} \left\| \mathbf{Q}_b \mathbf{v} \right\|^{m-2} \mathbf{Q}_b^t \mathbf{Q}_b \mathbf{v}.$$

It appears from the following equation

$$\text{(8)} \quad \mathbf{v}^t \nabla(\mathbf{v}) = m \sum_{b=1}^{B} \|\mathbf{Q}_b \mathbf{v}\|^m = m \Psi(\mathbf{v})$$

that the gradient $\nabla(\mathbf{v})$ is different from $\mathbf{0}$ for any point $\mathbf{v}$ such that $\Psi(\mathbf{v}) > 0$. We consider the Lagrangian function $F(\mathbf{v}, \lambda) = \Psi(\mathbf{v}) - \frac{\lambda}{2}(\mathbf{v}^t \mathbf{v} - 1)$ related to optimization problem (6). The optimal solution satisfies the stationary equation

$$\text{(9)} \quad \frac{\partial F}{\partial \mathbf{v}}(\mathbf{v}) = \nabla(\mathbf{v}) - \lambda \mathbf{v} = \mathbf{0}.$$

Using (8) and the constraint $\|\mathbf{v}\| = 1$, (9) can also be written as:

$$\text{(10)} \quad \frac{\partial F}{\partial \mathbf{v}}(\mathbf{v}) = \nabla(\mathbf{v}) - m \Psi(\mathbf{v}) \mathbf{v} = \mathbf{0}.$$

We denote a unit norm solution of the stationary equation (9) by

$$\text{(11)} \quad \mathbf{v}^* = \frac{\nabla(\mathbf{v}^*)}{\|\nabla(\mathbf{v}^*)\|}.$$

The globally optimal solution $\mathbf{v}^*$ of (6) is the solution of (10) maximizing $\Psi(\mathbf{v}^*)$. Using Algorithm 2 given in Appendix 1, we obtain Algorithm 1 (R-CPCA) for optimization problem (6). This algorithm guarantees a solution of the stationary equation (10), but not necessarily the globally optimal one. However, from our practical experience, the globally optimal solution is almost always reached whatever the randomly initialization point.

---

**Input:** $\mathbf{Q}_1, \ldots \mathbf{Q}_B, m, \varepsilon, \mathbf{v}_{B+1}^0$ such that $\Psi(\mathbf{v}_{B+1}^0) > 0$.
**Output:** $\mathbf{v}_{B+1}^s$ (approximate solution of (6)).
**Repeat** ($s = 0, 1, 2, \ldots$):

$$\text{(12)} \quad \mathbf{v}_{B+1}^{s+1} = f(\mathbf{v}_{B+1}^s) = \frac{\sum_{b=1}^{B} \|\mathbf{Q}_b \mathbf{v}_{B+1}^s\|^{m-2} \mathbf{Q}_b^t \mathbf{Q}_b \mathbf{v}_{B+1}^s}{\left\|\sum_{b=1}^{B} \|\mathbf{Q}_b \mathbf{v}_{B+1}^s\|^{m-2} \mathbf{Q}_b^t \mathbf{Q}_b \mathbf{v}_{B+1}^s\right\|}.$$

**Until** $\Psi(\mathbf{v}_{B+1}^{s+1}) - \Psi(\mathbf{v}_{B+1}^s) \leq \varepsilon$.

---

**ALGORITHM 1:** R-CPCA algorithm for optimization problem (6) ($m$ is any value $\geq 1$. $\varepsilon$ is a small positive constant that determines the desired accuracy of the algorithm)

In the field of multiblock component methods, the convergence we consider is "monotone convergence". A recurrence equation $\mathbf{v}^{s+1} = f(\mathbf{v}^s)$ is used to generate the sequence $\{\mathbf{v}^s\}$. Monotone convergence of the algorithm with respect to a criterion $\Psi$ means that the bounded sequence $\{\Psi(\mathbf{v}^s)\}$ is monotonically increasing and therefore convergent toward some limit $\Psi(\mathbf{v}^*)$. In practice, it will very often be noted that the limit point $\mathbf{v}^*$ is a fixed point of the recurrence equation: $\mathbf{v}^* = f(\mathbf{v}^*)$. But this property is usually not proven. In this paper we have been able to go further for a very large class of method. Proposition 1 below establishes three important points:

a) The algorithm is converging at the level of the $\Psi$ function: $\Psi(\mathbf{v}^s) \to \Psi(\mathbf{v}^*)$.

b) At convergence, the superblock weight-vector $\mathbf{v}^*$ is a fixed point of the recurrence equation.

c) If the number of fixed points of the recurrence equation is finite (mild condition), then $\mathbf{v}^s$ converges to a fixed point $\mathbf{v}^*$ of $f$.

We now consider the sequence $\{\mathbf{v}_{B+1}^s\}$ generated by the recurrence equation (12). We suppose that this sequence involves infinitely many distinct terms; otherwise the last given point satisfies $\mathbf{v}_{B+1}^s = f(\mathbf{v}_{B+1}^s)$, that is, $\mathbf{v}_{B+1}^s$ is a fixed point of $f$. Applying Proposition 2 of Appendix 1 yields the following proposition:

**Proposition 1**

Let $\{\mathbf{v}_{B+1}^s\}$ be a sequence of unit norm superblock weight-vectors generated by the recurrence equation (12). The following properties hold:

a) The bounded sequence $\Psi(\mathbf{v}_{B+1}^s) = \sum_{b=1}^{B} \|\mathbf{Q}_b \mathbf{v}_{B+1}^s\|^m$ is monotonically increasing and therefore converging for any value $m \geq 1$.

b) All accumulation points of the sequence $\{\mathbf{v}_{B+1}^s\}$ are fixed points of $f$.

c) $\Psi(\mathbf{v}_{B+1}^s) \to \Psi(\mathbf{v}_{B+1}^*)$, where $\mathbf{v}_{B+1}^*$ is a fixed point of $f$.

d) $\|\mathbf{v}_{B+1}^{s+1} - \mathbf{v}_{B+1}^s\|^2 \leq \dfrac{2\left(\Psi(\mathbf{v}_{B+1}^{s+1}) - \Psi(\mathbf{v}_{B+1}^s)\right)}{m\Psi(\mathbf{v}_{B+1}^0)}$.

e) The sequence $\{\mathbf{v}_{B+1}^s\}$ is asymptotically regular: $\lim\limits_{s \to \infty} \|\mathbf{v}_{B+1}^{s+1} - \mathbf{v}_{B+1}^s\| = 0$.

f) Either $\{\mathbf{v}_{B+1}^s\}$ converges or the accumulation points of $\{\mathbf{v}_{B+1}^s\}$ form a continuum.

**Comments**

1) A good starting point $\mathbf{v}_{B+1}^0$ is the eigenvector of $\sum_{b=1}^{B} \mathbf{Q}_b^t \mathbf{Q}_b$ corresponding to the largest eigenvalue.

2) From the Cauchy-Schwarz inequality, $m\Psi(\mathbf{v}_{B+1}^s) = (\mathbf{v}_{B+1}^s)^t \nabla(\mathbf{v}_{B+1}^s) \leq \|\nabla(\mathbf{v}_{B+1}^s)\|$. As $\Psi(\mathbf{v}_{B+1}^0)$ is strictly positive by construction, this implies $0 < m\Psi(\mathbf{v}_{B+1}^0) \leq \inf_s \|\nabla(\mathbf{v}_{B+1}^s)\| = \delta$. Then, Point (d) of Proposition 1 is deduced from Proposition 2 of Appendix 1 by noting that

$$(13) \quad \|\mathbf{v}_{B+1}^{s+1} - \mathbf{v}_{B+1}^s\|^2 \leq \frac{2\left(\Psi(\mathbf{v}_{B+1}^{s+1}) - \Psi(\mathbf{v}_{B+1}^s)\right)}{\delta} \leq \frac{2\left(\Psi(\mathbf{v}_{B+1}^{s+1}) - \Psi(\mathbf{v}_{B+1}^s)\right)}{m\Psi(\mathbf{v}_{B+1}^0)}.$$

At convergence of Algorithm 1, $\|\mathbf{v}_{B+1}^{s+1} - \mathbf{v}_{B+1}^s\| \leq 2\varepsilon / m\Psi(\mathbf{v}_{B+1}^0)$. When this inequality is satisfied, the solution $\mathbf{v}_{B+1}^s$ is considered as a fixed point of the recurrence equation (12).

3) When $m = 2$, Algorithm 1 is similar to the power iteration method and converges to the globally optimal solution of (6), i.e. the eigenvector of $\sum_{b=1}^{B} \mathbf{Q}_b^t \mathbf{Q}_b$ corresponding to the largest eigenvalue. In that case, many known multiblock component methods are recovered (see Section 3).

4) At convergence, equation (12) becomes

$$(14) \quad \mathbf{v}_{B+1}^* = \sum_{b=1}^{B} \|\mathbf{Q}_b \mathbf{v}_{B+1}^*\|^{m-2} \mathbf{Q}_b^t \mathbf{Q}_b \mathbf{v}_{B+1}^* \Big/ \left\|\sum_{b=1}^{B} \|\mathbf{Q}_b \mathbf{v}_{B+1}^*\|^{m-2} \mathbf{Q}_b^t \mathbf{Q}_b \mathbf{v}_{B+1}^*\right\|.$$

Therefore this fixed point $\mathbf{v}_{B+1}^*$ of $f$ is solution of the equation

$$(15) \quad \sum_{b=1}^{B} \|\mathbf{Q}_b \mathbf{v}_{B+1}^*\|^{m-2} \mathbf{Q}_b^t \mathbf{Q}_b \mathbf{v}_{B+1}^* = \Psi(\mathbf{v}_{B+1}^*) \mathbf{v}_{B+1}^*$$

and consequently a solution of the stationary equation (10).

5) If we suppose that the number of fixed points of $f$ is finite, then we can deduce from Point (f) of the proposition that the sequence $\{\mathbf{v}_{B+1}^s\}$ is converging. For $m = 2$, this condition corresponds to the assumption that all eigenvalues of $\sum_{b=1}^{B} \mathbf{Q}_b^t \mathbf{Q}_b$ are different.

## 2.3 A monotone convergent algorithm for optimization problem (2)

The R-CPCA algorithm for optimization problem (2) is immediately obtained by expressing Algorithm 1 in term of the original data: $\mathbf{v}_{B+1}^s$ is replaced by $\mathbf{M}_{B+1}^{1/2} \mathbf{w}_{B+1}^s$, $\mathbf{Q}_b = \mathbf{P}_b^t \mathbf{P}_{B+1}$ by $\mathbf{M}_b^{-1/2} \mathbf{X}_b^t \mathbf{X}_{B+1} \mathbf{M}_{B+1}^{-1/2}$ and $\mathbf{Q}_b^t \mathbf{Q}_b$ by $\mathbf{M}_{B+1}^{-1/2} \mathbf{X}_{B+1}^t \mathbf{X}_b \mathbf{M}_b^{-1} \mathbf{X}_b^t \mathbf{X}_{B+1} \mathbf{M}_{B+1}^{-1/2}$. Selecting a specific value of $m$ and metrics $\mathbf{M}_1, ..., \mathbf{M}_B, \mathbf{M}_{B+1}$ related to block and superblock weight-vectors, this single algorithm gives a new and unified framework for many multiblock component methods. We deduce from (12) recurrence equations on the superblock weight-vectors $\mathbf{w}_{B+1}^s$ and on the superblock component $\mathbf{y}_{B+1}^s = \mathbf{X}_{B+1} \mathbf{w}_{B+1}^s$. Asymptotic regularity of the sequence $\{\mathbf{v}_{B+1}^s\}$ also implies asymptotic regularity of the sequences $\{\mathbf{w}_{B+1}^s\}$ and $\{\mathbf{y}_{B+1}^s\}$. At convergence of the algorithm, the superblock weight-vector $\mathbf{w}_{B+1}$ and the superblock component $\mathbf{y}_{B+1} = \mathbf{X}_{B+1}\mathbf{w}_{B+1}$ are fixed points of their respective recurrence equations and therefore solutions of the following stationary equations:

$$(16) \quad \mathbf{w}_{B+1} = \frac{\mathbf{M}_{B+1}^{-1}\mathbf{X}_{B+1}^t \sum_{b=1}^{B} \left\|\mathbf{M}_b^{-1/2}\mathbf{X}_b^t\mathbf{X}_{B+1}\mathbf{w}_{B+1}\right\|^{m-2} \mathbf{X}_b\mathbf{M}_b^{-1}\mathbf{X}_b^t\mathbf{X}_{B+1}\mathbf{w}_{B+1}}{\left\|\mathbf{M}_{B+1}^{-1/2}\mathbf{X}_{B+1}^t \sum_{b=1}^{B} \left\|\mathbf{M}_b^{-1/2}\mathbf{X}_b^t\mathbf{X}_{B+1}\mathbf{w}_{B+1}\right\|^{m-2} \mathbf{X}_b\mathbf{M}_b^{-1}\mathbf{X}_b^t\mathbf{X}_{B+1}\mathbf{w}_{B+1}\right\|},$$

$$(17) \quad \mathbf{y}_{B+1} = \frac{\mathbf{X}_{B+1}\mathbf{M}_{B+1}^{-1}\mathbf{X}_{B+1}^t \sum_{b=1}^{B} \left\|\mathbf{M}_b^{-1/2}\mathbf{X}_b^t\mathbf{y}_{B+1}\right\|^{m-2} \mathbf{X}_b\mathbf{M}_b^{-1}\mathbf{X}_b^t\mathbf{y}_{B+1}}{\left\|\mathbf{M}_{B+1}^{-1/2}\mathbf{X}_{B+1}^t \sum_{b=1}^{B} \left\|\mathbf{M}_b^{-1/2}\mathbf{X}_b^t\mathbf{y}_{B+1}\right\|^{m-2} \mathbf{X}_b\mathbf{M}_b^{-1}\mathbf{X}_b^t\mathbf{y}_{B+1}\right\|}.$$

Using (4), the block weight-vector $\mathbf{w}_b$ is related to the superblock weight-vector and component by the equation

$$(18) \quad \mathbf{w}_b = \mathbf{M}_b^{-1}\mathbf{X}_b^t\mathbf{X}_{B+1}\mathbf{w}_{B+1} / \left\|\mathbf{M}_b^{-1/2}\mathbf{X}_b^t\mathbf{X}_{B+1}\mathbf{w}_{B+1}\right\| = \mathbf{M}_b^{-1}\mathbf{X}_b^t\mathbf{y}_{B+1} / \left\|\mathbf{M}_b^{-1/2}\mathbf{X}_b^t\mathbf{y}_{B+1}\right\|.$$

Then, the block component $\mathbf{y}_b = \mathbf{X}_b\mathbf{w}_b$ is related to the superblock component by the equation

$$(19) \quad \mathbf{y}_b = \mathbf{X}_b\mathbf{w}_b = \mathbf{X}_b\mathbf{M}_b^{-1}\mathbf{X}_b^t\mathbf{y}_{B+1} / \left\|\mathbf{M}_b^{-1/2}\mathbf{X}_b^t\mathbf{y}_{B+1}\right\|.$$

The following relation is also useful:

$$(20) \quad \mathrm{cov}(\mathbf{y}_b, \mathbf{y}_{B+1}) = \frac{1}{n}\mathbf{y}_{B+1}^t\mathbf{y}_b = \frac{1}{n}\left\|\mathbf{M}_b^{-1/2}\mathbf{X}_b^t\mathbf{y}_{B+1}\right\|.$$

Conversely, using (19) and (20) in equation (17), the superblock component is related to the block components:

$$(21) \quad \mathbf{y}_{B+1} = \frac{\mathbf{X}_{B+1}\mathbf{M}_{B+1}^{-1}\mathbf{X}_{B+1}^t \sum_{b=1}^{B} \left(\mathrm{cov}(\mathbf{y}_b, \mathbf{y}_{B+1})\right)^{m-1} \mathbf{y}_b}{\left\|\mathbf{M}_{B+1}^{-1/2}\mathbf{X}_{B+1}^t \sum_{b=1}^{B} \left(\mathrm{cov}(\mathbf{y}_b, \mathbf{y}_{B+1})\right)^{m-1} \mathbf{y}_b\right\|}.$$

Note that the only case where the superblock component is proportional to the sum of the block components occurs when $m = 1$ and Mode B selected for the superblock.

At convergence, the contribution of the block component $\mathbf{y}_b$ to the superblock component $\mathbf{y}_{B+1}$ may be defined by using the following indicator:

$$(22) \quad c_b = \left(\text{cov}(\mathbf{y}_b, \mathbf{y}_{B+1})\right)^m / \sum_{b=1}^{B} \left(\text{cov}(\mathbf{y}_b, \mathbf{y}_{B+1})\right)^m.$$

We can conclude that the contributions of the most representative blocks of the superblock are even greater than $m$ is large: when $m \to \infty$, $c_b \to 1$ for the most contributive block to the solution and $c_b \to 0$ for the others.

A regularized consensus PCA method is defined by the choices of $m$ and of the metrics $\mathbf{M}_1, ..., \mathbf{M}_B, \mathbf{M}_{B+1}$. Stationary equation (17) is the *signature* of a regularized consensus PCA method as we can deduce from this equation the related optimization problem. Equation (17) is a generalization of the equation (17) given in Smilde, Westerhuis and de Jong (2003). Let us give an illustration. We consider the hierarchical PCA method described in Smilde, Westerhuis, and de Jong (2003). The stationary equation corresponding to this method, written in our notation, is $\mathbf{y}_{B+1} \propto \sum_{b=1}^{B} \|\mathbf{X}_b^t \mathbf{y}_{B+1}\|^2 \mathbf{X}_b \mathbf{X}_b^t \mathbf{y}_{B+1}$ where $\propto$ means that the left term is proportional to the right term. A regularized consensus PCA method sharing this stationary equation is obtained for $m = 4$, $\mathbf{M}_1 = ... = \mathbf{M}_B = \mathbf{I}$ and $\mathbf{M}_{B+1} = (1/n)\mathbf{X}_{B+1}^t \mathbf{X}_{B+1}$. Therefore, we may conclude that this hierarchical PCA method is similar to the regularized consensus PCA method corresponding to the following optimization problem:

$$(23) \quad \underset{\mathbf{w}_1, ..., \mathbf{w}_{B+1}}{\text{Maximize}} \sum_{b=1}^{B} \left(\text{cov}(\mathbf{X}_b \mathbf{w}_b, \mathbf{X}_{B+1} \mathbf{w}_{B+1})\right)^4$$
$$\text{s.t. } \mathbf{w}_b^t \mathbf{w}_b = 1, \ b = 1, ..., B \text{ and } \text{var}(\mathbf{X}_{B+1} \mathbf{w}_{B+1}) = 1.$$

This property of hierarchical PCA has been previously established in Hanafi, Kohler and Qannari (2010).

## 2.4 Simplifications occurring when Mode A or B are selected

### When Mode A is selected for all blocks

In this special case, using (19) and (20) with $\mathbf{M}_b = \mathbf{I}$, we obtain $\mathbf{y}_b \mathbf{y}_b^t \mathbf{y}_{B+1} = \mathbf{X}_b \mathbf{X}_b^t \mathbf{y}_{B+1}$ and therefore, noting $\mathbf{Y} = [\mathbf{y}_1, ..., \mathbf{y}_B]$, we deduce:

$$(24) \quad \mathbf{YY}^t \mathbf{y}_{B+1} = \sum_{b=1}^{B} \mathbf{y}_b \mathbf{y}_b^t \mathbf{y}_{B+1} = \mathbf{X}_{B+1} \mathbf{X}_{B+1}^t \mathbf{y}_{B+1}.$$

Consequently, if the superblock component is the first principal component of the superblock $\mathbf{X}_{B+1}$, then it is also the first principal component of the block components $\mathbf{y}_1, ..., \mathbf{y}_B$. Such situation is encountered in Consensus PCA and similar methods (see Section 3 below and methods 5 and 6 in Table 1).

### When Mode B is selected for the superblock

When Mode B is selected for the superblock $\mathbf{X}_{B+1}$, equations (17) and (21) can be simplified. As the superblock $\mathbf{X}_{B+1}$ is the concatenation of all blocks, for $\mathbf{M}_{B+1} = (1/n)\mathbf{X}_{B+1}^t \mathbf{X}_{B+1}$, equations (17) and (21) become respectively

$$(25) \quad \mathbf{y}_{B+1} = \frac{\sum_{b=1}^{B} \left\| \mathbf{M}_b^{-1/2} \mathbf{X}_b^t \mathbf{y}_{B+1} \right\|^{m-2} \mathbf{X}_b \mathbf{M}_b^{-1} \mathbf{X}_b^t \mathbf{y}_{B+1}}{\mathrm{var}\left( \sum_{b=1}^{B} \left\| \mathbf{M}_b^{-1/2} \mathbf{X}_b^t \mathbf{y}_{B+1} \right\|^{m-2} \mathbf{X}_b \mathbf{M}_b^{-1} \mathbf{X}_b^t \mathbf{y}_{B+1} \right)^{1/2}}$$

and

$$(26) \quad \mathbf{y}_{B+1} = \frac{\sum_{b=1}^{B} \left( \mathrm{cov}(\mathbf{y}_b, \mathbf{y}_{B+1}) \right)^{m-1} \mathbf{y}_b}{\mathrm{var}\left( \sum_{b=1}^{B} \left( \mathrm{cov}(\mathbf{y}_b, \mathbf{y}_{B+1}) \right)^{m-1} \mathbf{y}_b \right)^{1/2}}.$$

Note that when Mode B is selected for the superblock, this superblock does not appear anymore in the stationary equation (25). In fact, using a similar development as the one used for (2), it can be shown that (25) is also the stationary equation of (1). Optimization problem (1) is apparently less constrained than (2) as $\mathbf{y}_{B+1}$ is not constrained to belong to the superblock space. But, due to (25), it will be the case at the optimum. Therefore, we may conclude that, when Mode B is selected for the superblock, optimization problems (1) and (2) are equivalent. Note that equation (26) remains also valid for problem (1).

Other properties are worth to be mentioned.

*For $m = 1$*

In this situation, the superblock component, solution of (25), is proportional to the sum of the block components. In fact, for $m = 1$ equation (26) simplifies to

$$(27) \quad \mathbf{y}_{B+1} = \sum_{b=1}^{B} \mathbf{y}_b \Big/ \mathrm{var}\Big(\sum_{b=1}^{B} \mathbf{y}_b\Big)^{1/2}.$$

*For $m = 1$ and Mode B selected for all blocks*

When $m = 1$ and Mode B is selected for all blocks, (1) becomes

$$(28) \quad \underset{\mathbf{w}_1,\ldots,\mathbf{w}_B, \mathbf{y}_{B+1}}{\mathrm{Maximize}} \sum_{b=1}^{B} \mathrm{cov}(\mathbf{X}_b \mathbf{w}_b, \mathbf{y}_{B+1})$$
$$\text{s.t. } \mathrm{var}(\mathbf{X}_b \mathbf{w}_b) = 1, \ b = 1,\ldots,B, \text{ and } \mathrm{var}(\mathbf{y}_{B+1}) = 1.$$

As a superblock component, solution of (28), verifies (27), the criterion in (28) can be expressed in term of block components only. Using (27) and the normalization constraints, we obtain:

$$\sum_{b=1}^{B} \mathrm{cov}(\mathbf{X}_b \mathbf{w}_b, \mathbf{y}_{B+1}) = \mathrm{cov}\left(\sum_{b=1}^{B} \mathbf{X}_b \mathbf{w}_b, \frac{\sum_{b=1}^{B} \mathbf{X}_b \mathbf{w}_b}{\mathrm{var}(\sum_{b=1}^{B} \mathbf{X}_b \mathbf{w}_b)^{1/2}}\right)$$

$$= \mathrm{var}\left(\sum_{b=1}^{B} \mathbf{X}_b \mathbf{w}_b\right)^{1/2} = \left(\sum_{b,c=1}^{B} \mathrm{cor}(\mathbf{X}_b \mathbf{w}_b, \mathbf{X}_c \mathbf{w}_c)\right)^{1/2}. \quad (29)$$

We conclude that (28) is equivalent to the following optimization problem:

$$(30) \quad \underset{\mathbf{w}_1,\ldots,\mathbf{w}_B}{\text{Maximize}} \sum_{b,c=1}^{B} \mathrm{cor}(\mathbf{X}_b \mathbf{w}_b, \mathbf{X}_c \mathbf{w}_c)$$

$$\text{s. t. } \mathrm{var}(\mathbf{X}_b \mathbf{w}_b) = 1, \ b = 1,\ldots B.$$

This problem (30) corresponds exactly to the SUMCOR method proposed by Horst (1961a,b, 1965).

*For m = 2*

In this special case, the superblock component is the standardized first principal component of the block components $\mathbf{y}_1,\ldots,\mathbf{y}_B$. Indeed, when Mode B is selected for the superblock, using (20) in equation (26), we obtain:

$$(31) \quad \mathbf{y}_{B+1} \propto \mathbf{Y}\mathbf{Y}^t \mathbf{y}_{B+1}.$$

*When the superblock $\mathbf{X}_{B+1}$ is not full rank*

Equation (25) has been established under the assumption that the superblock $\mathbf{X}_{B+1}$ is full rank. However, as it is also the stationary equation of optimization problem (1), it remains valid when $\mathbf{X}_{B+1}$ is not full rank. Let $\mathbf{y}_{B+1}$ be a solution of (25). We deduce from (25) that some superblock weight-vector $\mathbf{w}_{B+1}$ exists such that $\mathbf{y}_{B+1} = \mathbf{X}_{B+1}\mathbf{w}_{B+1}$. But this $\mathbf{w}_{B+1}$ is not unique because $\mathbf{X}_{B+1}$ is not full rank.

*Conclusion*

We conclude that, when Mode B is selected for the superblock, it is preferable to interpret the superblock component $\mathbf{y}_{B+1}$ by considering $\mathrm{cor}(\mathbf{y}_b, \mathbf{y}_{B+1})$ and $\mathrm{cor}(\mathbf{x}_{B+1,j}, \mathbf{y}_{B+1})$, where $\mathbf{x}_{B+1,j}$ is a column of $\mathbf{X}_{B+1}$. In fact, the superblock weight-vector $\mathbf{w}_{B+1}$ is difficult to interpret because it is sensible to a possible multicolinearity when $\mathbf{X}_{B+1}$ is full rank and is not uniquely determined when $\mathbf{X}_{B+1}$ is not full rank.

*When Mode B is selected for a block*

When Mode B is selected for a block $b$, the corresponding term appearing in equation (17) becomes:

$$(32) \quad \left\| \left( \frac{1}{n} \mathbf{X}_b^t \mathbf{X}_b \right)^{-1/2} \mathbf{X}_b^t \mathbf{y}_{B+1} \right\|^{m-2} \mathbf{X}_b \left( \frac{1}{n} \mathbf{X}_b^t \mathbf{X}_b \right)^{-1} \mathbf{X}_b^t \mathbf{y}_{B+1}$$

$$= n^{\frac{m}{2}} \left\| \mathbf{X}_b \left( \mathbf{X}_b^t \mathbf{X}_b \right)^{-1} \mathbf{X}_b^t \mathbf{y}_{B+1} \right\|^{m-2} \mathbf{X}_b \left( \mathbf{X}_b^t \mathbf{X}_b \right)^{-1} \mathbf{X}_b^t \mathbf{y}_{B+1}.$$

Using generalized inverse, this term is still meaningful for a not full rank block as the projection operator $\mathbf{X}_b \left( \mathbf{X}_b^t \mathbf{X}_b \right)^{-} \mathbf{X}_b^t$ is invariant to the choice of any generalized inverse $\left( \mathbf{X}_b^t \mathbf{X}_b \right)^{-}$. This result shows how regularized consensus PCA can be extended to situations where Mode B is selected for a not full rank block: first, replace $\mathbf{X}_b \mathbf{w}_b$ by $\mathbf{y}_b$ in the criterion used in problem (2); secondly, impose to the block component $\mathbf{y}_b$ to be standardized and to belong to the $\mathbf{X}_b$ space; and thirdly, compute the superblock component $\mathbf{y}_{B+1}$ by solving equation (17) with $\left( \mathbf{X}_b^t \mathbf{X}_b \right)^{-1}$ replaced by the Moore-Penrose inverse $\left( \mathbf{X}_b^t \mathbf{X}_b \right)^{+}$. The block component $\mathbf{y}_b$ is then deduced from (19):

$$(33) \quad \mathbf{y}_b = \frac{\mathbf{X}_b \left( \mathbf{X}_b^t \mathbf{X}_b \right)^{+} \mathbf{X}_b^t \mathbf{y}_{B+1}}{\mathrm{var}\left( \mathbf{X}_b \left( \mathbf{X}_b^t \mathbf{X}_b \right)^{+} \mathbf{X}_b^t \mathbf{y}_{B+1} \right)^{1/2}}.$$

## 3. Regularized Consensus PCA: a new framework for multiblock component methods

In this section, we discuss several special cases of R-CPCA with Modes A or B selected for the blocks and the superblock. They are summarized in Table 1. The superblock stationary equation of an R-CPCA method allows obtaining the superblock component by using Algorithm 1. Then, each block component is obtained by regression of the superblock component on the block: OLS regression if Mode B is selected for the block, one-component PLS regression if Mode A is selected.

**Table 1:** Special cases of Regularized Consensus PCA methods

| Method | Mode for Blocks | Mode for Superblock | Optimization problem | Stationary equation for the superblock component |
|---|---|---|---|---|
| | | | **m = 1** | |
| 1 | A | A | $\underset{\mathbf{w}_1,\ldots,\mathbf{w}_{B+1}}{\text{Maximize}} \sum_{b=1}^{B} \text{cov}(\mathbf{X}_b\mathbf{w}_b, \mathbf{X}_{B+1}\mathbf{w}_{B+1})$ s.t. $\|\mathbf{w}_b\| = 1, b = 1,\ldots, B+1$ | $\mathbf{y}_{B+1} \propto \mathbf{X}_{B+1}\mathbf{X}_{B+1}^t \sum_{b=1}^{B} \|\mathbf{X}_b^t \mathbf{y}_{B+1}\|^{-1} \mathbf{X}_b \mathbf{X}_b^t \mathbf{y}_{B+1}$ |
| 2 | A | B | $\underset{\mathbf{w}_1,\ldots,\mathbf{w}_{B+1}}{\text{Maximize}} \sum_{b=1}^{B} \text{cor}(\mathbf{X}_b\mathbf{w}_b, \mathbf{X}_{B+1}\mathbf{w}_{B+1}) \text{var}(\mathbf{X}_b\mathbf{w}_b)^{1/2}$ s.t. $\|\mathbf{w}_b\| = 1, b = 1,\ldots, B, \text{var}(\mathbf{X}_{B+1}\mathbf{w}_{B+1}) = 1$ | $\mathbf{y}_{B+1} \propto \sum_{b=1}^{B} \|\mathbf{X}_b^t \mathbf{y}_{B+1}\|^{-1} \mathbf{X}_b \mathbf{X}_b^t \mathbf{y}_{B+1}$ |
| 3 | B | A | $\underset{\mathbf{w}_1,\ldots,\mathbf{w}_{B+1}}{\text{Maximize}} \sum_{b=1}^{B} \text{cor}(\mathbf{X}_b\mathbf{w}_b, \mathbf{X}_{B+1}\mathbf{w}_{B+1}) \text{var}(\mathbf{X}_{B+1}\mathbf{w}_{B+1})^{1/2}$ s.t. $\text{var}(\mathbf{X}_b\mathbf{w}_b) = 1, b = 1,\ldots, B, \|\mathbf{w}_{B+1}\| = 1$ | $\mathbf{y}_{B+1} \propto \mathbf{X}_{B+1}\mathbf{X}_{B+1}^t \sum_{b=1}^{B} \dfrac{\mathbf{X}_b(\mathbf{X}_b^t\mathbf{X}_b)^{-1}\mathbf{X}_b^t\mathbf{y}_{B+1}}{\|\mathbf{X}_b(\mathbf{X}_b^t\mathbf{X}_b)^{-1}\mathbf{X}_b^t\mathbf{y}_{B+1}\|}$ |
| 4 | B | B | $\underset{\mathbf{w}_1,\ldots,\mathbf{w}_{B+1}}{\text{Maximize}} \sum_{b=1}^{B} \text{cor}(\mathbf{X}_b\mathbf{w}_b, \mathbf{X}_{B+1}\mathbf{w}_{B+1})$ s.t. $\text{var}(\mathbf{X}_b\mathbf{w}_b) = 1, b = 1,\ldots, B+1$ | $\mathbf{y}_{B+1} \propto \sum_{b=1}^{B} \dfrac{\mathbf{X}_b(\mathbf{X}_b^t\mathbf{X}_b)^{-1}\mathbf{X}_b^t\mathbf{y}_{B+1}}{\|\mathbf{X}_b(\mathbf{X}_b^t\mathbf{X}_b)^{-1}\mathbf{X}_b^t\mathbf{y}_{B+1}\|}$ |
| | | | **m = 2** | |
| 5 | A | A | $\underset{\mathbf{w}_1,\ldots,\mathbf{w}_{B+1}}{\text{Maximize}} \sum_{b=1}^{B} \text{cov}^2(\mathbf{X}_b\mathbf{w}_b, \mathbf{X}_{B+1}\mathbf{w}_{B+1})$ s.t. $\|\mathbf{w}_b\| = 1, b = 1,\ldots, B+1$ | $\mathbf{y}_{B+1} \propto \mathbf{X}_{B+1}\mathbf{X}_{B+1}^t \mathbf{y}_{B+1}$ |
| 6 | A | B | $\underset{\mathbf{w}_1,\ldots,\mathbf{w}_{B+1}}{\text{Maximize}} \sum_{b=1}^{B} \text{cor}^2(\mathbf{X}_b\mathbf{w}_b, \mathbf{X}_{B+1}\mathbf{w}_{B+1}) \text{var}(\mathbf{X}_b\mathbf{w}_b)$ s.t. $\|\mathbf{w}_b\| = 1, b = 1,\ldots, B, \text{var}(\mathbf{X}_{B+1}\mathbf{w}_{B+1}) = 1$ | |
| 7 | B | A | $\underset{\mathbf{w}_1,\ldots,\mathbf{w}_{B+1}}{\text{Maximize}} \sum_{b=1}^{B} \text{cor}^2(\mathbf{X}_b\mathbf{w}_b, \mathbf{X}_{B+1}\mathbf{w}_{B+1}) \text{var}(\mathbf{X}_{B+1}\mathbf{w}_{B+1})$ s.t. $\text{var}(\mathbf{X}_b\mathbf{w}_b) = 1, b = 1,\ldots, B, \|\mathbf{w}_{B+1}\| = 1$ | $\mathbf{y}_{B+1} \propto \mathbf{X}_{B+1}\mathbf{X}_{B+1}^t \sum_{b=1}^{B} \mathbf{X}_b(\mathbf{X}_b^t\mathbf{X}_b)^{-1}\mathbf{X}_b^t\mathbf{y}_{B+1}$ |
| 8 | B | B | $\underset{\mathbf{w}_1,\ldots,\mathbf{w}_{B+1}}{\text{Maximize}} \sum_{b=1}^{B} \text{cor}^2(\mathbf{X}_b\mathbf{w}_b, \mathbf{X}_{B+1}\mathbf{w}_{B+1})$ s.t. $\text{var}(\mathbf{X}_b\mathbf{w}_b) = 1, b = 1,\ldots, B+1$ | $\mathbf{y}_{B+1} \propto \sum_{b=1}^{B} \mathbf{X}_b(\mathbf{X}_b^t\mathbf{X}_b)^{-1}\mathbf{X}_b^t\mathbf{y}_{B+1}$ |
| | | | **m = 4** | |
| 9 | A | A | $\underset{\mathbf{w}_1,\ldots,\mathbf{w}_{B+1}}{\text{Maximize}} \sum_{b=1}^{B} \text{cov}^4(\mathbf{X}_b\mathbf{w}_b, \mathbf{X}_{B+1}\mathbf{w}_{B+1})$ s.t. $\|\mathbf{w}_b\| = 1, b = 1,\ldots, B+1$ | $\mathbf{y}_{B+1} \propto \mathbf{X}_{B+1}\mathbf{X}_{B+1}^t \sum_{b=1}^{B} \|\mathbf{X}_b^t\mathbf{y}_{B+1}\|^2 \mathbf{X}_b\mathbf{X}_b^t\mathbf{y}_{B+1}$ |
| 10 | A | B | $\underset{\mathbf{w}_1,\ldots,\mathbf{w}_{B+1}}{\text{Maximize}} \sum_{b=1}^{B} \text{cor}^4(\mathbf{X}_b\mathbf{w}_b, \mathbf{X}_{B+1}\mathbf{w}_{B+1}) \text{var}^2(\mathbf{X}_b\mathbf{w}_b)$ s.t. $\|\mathbf{w}_b\| = 1, b = 1,\ldots, B, \text{var}(\mathbf{X}_{B+1}\mathbf{w}_{B+1}) = 1$ | $\mathbf{y}_{B+1} \propto \sum_{b=1}^{B} \|\mathbf{X}_b^t\mathbf{y}_{B+1}\|^2 \mathbf{X}_b\mathbf{X}_b^t\mathbf{y}_{B+1}$ |

**Table 1 (continued)**

| Method | Mode for Blocks | Mode for Superblock | Block component | Superblock component in term of block components |
|---|---|---|---|---|
| | | | $m = 1$ | |
| 1 | A | A | $\mathbf{y}_b = \mathbf{X}_b \mathbf{X}_b^t \mathbf{y}_{B+1} / \|\mathbf{X}_b^t \mathbf{y}_{B+1}\|$ | $\mathbf{y}_{B+1} \propto \mathbf{X}_{B+1} \mathbf{X}_{B+1}^t \sum_{b=1}^{B} \mathbf{y}_b$ |
| 2 | A | B | $\mathbf{y}_b = \mathbf{X}_b \mathbf{X}_b^t \mathbf{y}_{B+1} / \|\mathbf{X}_b^t \mathbf{y}_{B+1}\|$ | $\mathbf{y}_{B+1} \propto \sum_{b=1}^{B} \mathbf{y}_b$ |
| 3 | B | A | $\mathbf{y}_b = \dfrac{\mathbf{X}_b (\mathbf{X}_b^t \mathbf{X}_b)^{-1} \mathbf{X}_b^t \mathbf{y}_{B+1}}{\mathrm{var}\left(\mathbf{X}_b (\mathbf{X}_b^t \mathbf{X}_b)^{-1} \mathbf{X}_b^t \mathbf{y}_{B+1}\right)^{1/2}}$ | $\mathbf{y}_{B+1} \propto \mathbf{X}_{B+1} \mathbf{X}_{B+1}^t \sum_{b=1}^{B} \mathbf{y}_b$ |
| 4 | B | B | $\mathbf{y}_b = \dfrac{\mathbf{X}_b (\mathbf{X}_b^t \mathbf{X}_b)^{-1} \mathbf{X}_b^t \mathbf{y}_{B+1}}{\mathrm{var}\left(\mathbf{X}_b (\mathbf{X}_b^t \mathbf{X}_b)^{-1} \mathbf{X}_b^t \mathbf{y}_{B+1}\right)^{1/2}}$ | $\mathbf{y}_{B+1} \propto \sum_{b=1}^{B} \mathbf{y}_b$ |
| | | | $m = 2$ | |
| 5 | A | A | $\mathbf{y}_b = \mathbf{X}_b \mathbf{X}_b^t \mathbf{y}_{B+1} / \|\mathbf{X}_b^t \mathbf{y}_{B+1}\|$ | $\mathbf{y}_{B+1} \propto \sum_{b=1}^{B} \mathrm{cov}(\mathbf{y}_b, \mathbf{y}_{B+1}) \mathbf{y}_b$ |
| 6 | A | B | | |
| 7 | B | A | $\mathbf{y}_b = \dfrac{\mathbf{X}_b (\mathbf{X}_b^t \mathbf{X}_b)^{-1} \mathbf{X}_b^t \mathbf{y}_{B+1}}{\mathrm{var}\left(\mathbf{X}_b (\mathbf{X}_b^t \mathbf{X}_b)^{-1} \mathbf{X}_b^t \mathbf{y}_{B+1}\right)^{1/2}}$ | $\mathbf{y}_{B+1} \propto \mathbf{X}_{B+1} \mathbf{X}_{B+1}^t \sum_{b=1}^{B} \mathrm{cov}(\mathbf{y}_b, \mathbf{y}_{B+1}) \mathbf{y}_b$ |
| 8 | B | B | $\mathbf{y}_b = \dfrac{\mathbf{X}_b (\mathbf{X}_b^t \mathbf{X}_b)^{-1} \mathbf{X}_b^t \mathbf{y}_{B+1}}{\mathrm{var}\left(\mathbf{X}_b (\mathbf{X}_b^t \mathbf{X}_b)^{-1} \mathbf{X}_b^t \mathbf{y}_{B+1}\right)^{1/2}}$ | $\mathbf{y}_{B+1} \propto \sum_{b=1}^{B} \mathrm{cor}(\mathbf{y}_b, \mathbf{y}_{B+1}) \mathbf{y}_b$ |
| | | | $m = 4$ | |
| 9 | A | A | $\mathbf{y}_b = \mathbf{X}_b \mathbf{X}_b^t \mathbf{y}_{B+1} / \|\mathbf{X}_b^t \mathbf{y}_{B+1}\|$ | $\mathbf{y}_{B+1} \propto \mathbf{X}_{B+1} \mathbf{X}_{B+1}^t \sum_{b=1}^{B} \left(\mathrm{cov}(\mathbf{y}_b, \mathbf{y}_{B+1})\right)^3 \mathbf{y}_b$ |
| 10 | A | B | $\mathbf{y}_b = \mathbf{X}_b \mathbf{X}_b^t \mathbf{y}_{B+1} / \|\mathbf{X}_b^t \mathbf{y}_{B+1}\|$ | $\mathbf{y}_{B+1} \propto \sum_{b=1}^{B} \left(\mathrm{cov}(\mathbf{y}_b, \mathbf{y}_{B+1})\right)^3 \mathbf{y}_b$ |

**Discussion**

When Mode A is used for all blocks and for the superblock, optimization problem (2) becomes

(34)
$$\underset{\mathbf{w}_1,...,\mathbf{w}_{B+1}}{\text{Maximize}} \sum_{b=1}^{B} (\text{cov}(\mathbf{X}_b\mathbf{w}_b, \mathbf{X}_{B+1}\mathbf{w}_{B+1}))^m$$
$$\text{s.t. } \mathbf{w}_b^t \mathbf{w}_b = 1, \ b = 1,...,B+1.$$

When optimization problem (34) is considered with $m = 2$, the following methods are recovered (only the normalization used for the block and superblock components varies from one method to another): Covariance criterion (Carroll, 1968b), Split principal component (Lohmöller, 1989; more precisely see p. 132, Table 3.12, row 6 which corresponds to applying the PLS path modeling algorithm to the consensus PCA model with the path weighting scheme and Mode A for all blocks and superblock), Multiple factor analysis (Escofier, Pagès, 1994), Multiple Co-Inertia Analysis (Chessel, Hanafi, 1996), Consensus PCA (Westerhuis, Kourti, MacGregor, 1998), CPCA-W (Smilde, Westerhuis, de Jong, 2001), one-dimension MAXBET A (Hanafi and Kiers, 2006) . In these methods, the superblock component $\mathbf{y}_{B+1} = \mathbf{X}_{B+1}\mathbf{w}_{B+1}$ is the first principal component of the superblock and at the same time that of $\mathbf{Y} = [\mathbf{y}_1,...,\mathbf{y}_B]$.

When Mode B is used for all blocks and for the superblock, optimization problem (2) becomes

(35)
$$\underset{\mathbf{w}_1,...,\mathbf{w}_{B+1}}{\text{Maximize}} \sum_{b=1}^{B} \left(\text{cor}(\mathbf{X}_b\mathbf{w}_b, \mathbf{X}_{B+1}\mathbf{w}_{B+1})\right)^m$$
$$\text{s.t. } \text{var}(\mathbf{X}_b\mathbf{w}_b) = 1, \ b = 1,...,B+1.$$

When optimization problem (35) is considered with $m = 2$, the MAXVAR method (Horst, 1961b, 1965) and the generalized canonical correlation analysis (Carroll, 1968) are found again. For $m = 1$, as discussed above, the Horst's SUMCOR method is also obtained.

When Mode A is used for all blocks and Mode B for the superblock, optimization problem (2) becomes

(36)
$$\underset{\mathbf{w}_1,...,\mathbf{w}_{B+1}}{\text{Maximize}} \sum_{b=1}^{B} \left(\text{cor}(\mathbf{X}_b\mathbf{w}_b, \mathbf{X}_{B+1}\mathbf{w}_{B+1}) \text{var}(\mathbf{X}_b\mathbf{w}_b)^{1/2}\right)^m$$
$$\text{s.t. } \mathbf{w}_b^t \mathbf{w}_b = 1, \ b = 1,...,B, \ \text{var}(\mathbf{X}_{B+1}\mathbf{w}_{B+1}) = 1.$$

This optimization problem is considered when the user wants to favor the blocks compared to the superblock. The objective of (36) is to find block components simultaneously well explaining their own block and well correlated to the superblock component. We remind that redundancy analysis of $\mathbf{X}_b$ with respect to $\mathbf{X}_{B+1}$ corresponds to the following optimization problem

$$\text{(37)} \quad \underset{\mathbf{w}_b, \mathbf{w}_{B+1}}{\text{Maximize}} \ \text{cor}(\mathbf{X}_b\mathbf{w}_b, \mathbf{X}_{B+1}\mathbf{w}_{B+1}) \, \text{var}(\mathbf{X}_b\mathbf{w}_b)^{1/2}$$
$$\text{s.t.} \ \|\mathbf{w}_b\| = 1, \text{var}(\mathbf{X}_{B+1}\mathbf{w}_{B+1}) = 1.$$

Therefore, optimization problem (36) is a generalized redundancy analysis of $\mathbf{X}_1, ..., \mathbf{X}_B$ with respect to $\mathbf{X}_{B+1}$. For $m = 2$, optimization problems (34) and (36) have the same solution up to the normalization of the superblock component. For $m = 4$, (36) yields hierarchical PCA (Smilde, Westerhuis, de Jong, 2003).

When mode B is used for all blocks and Mode A for the superblock, optimization problem (2) becomes

$$\text{(38)} \quad \underset{\mathbf{w}_1, ..., \mathbf{w}_{B+1}}{\text{Maximize}} \ \sum_{b=1}^{B} \left( \text{cor}(\mathbf{X}_b\mathbf{w}_b, \mathbf{X}_{B+1}\mathbf{w}_{B+1}) \, \text{var}(\mathbf{X}_{B+1}\mathbf{w}_{B+1})^{1/2} \right)^m$$
$$\text{s.t.} \ \text{var}(\mathbf{X}_b\mathbf{w}_b) = 1, \ b = 1, ..., B, \ \mathbf{w}_{B+1}^t \mathbf{w}_{B+1} = 1.$$

This optimization problem is considered when the user wants to favor the superblock compared to the blocks. The objective of (38) is to find a superblock component simultaneously well explaining the whole set of blocks and well correlated to the various block components. Optimization problem (38) is a generalized redundancy analysis of $\mathbf{X}_{B+1}$ with respect to $\mathbf{X}_1, ..., \mathbf{X}_B$.

We give in Table 2 a guideline to select the modes for blocks and superblock according to the objective of the user.

**Table 2:** Guideline for selecting Modes A or B for blocks and superblock

| Mode for | | Generalization of: | Objective: |
|---|---|---|---|
| Blocks | Superblock | | |
| A | A | Tucker's inter-battery factor analysis | Compromise between block and superblock components well explaining their own blocks and as correlated as possible. |
| A | B | Redundancy analysis of a block with respect to the superblock | Compromise between block components well explaining their own blocks and as correlated as possible to the superblock component. |
| B | A | Redundancy analysis of the superblock with respect to a block | Compromise between a superblock component well explaining the superblock and as correlated as possible to the block components. |
| B | B | Canonical correlation analysis | Block and superblock components as correlated as possible. |

# 4. Using deflation for searching higher order components

The deflation procedure consists in replacing a data matrix $\mathbf{X}$ by a matrix $\mathbf{E}$ where $\mathbf{E}$ is the residual matrix in the regression of $\mathbf{X}$ on some column vector $\mathbf{q}$: $\mathbf{E} = \mathbf{X} - \mathbf{q}\mathbf{q}^t\mathbf{X}/\mathbf{q}^t\mathbf{q}$. The obtained property is of course that any linear combination of the columns of $\mathbf{E}$ is orthogonal to $\mathbf{q}$. Furthermore, if $\mathbf{q}$ belongs to the $\mathbf{X}-$space ($\mathbf{q} = \mathbf{Xw}$), then the set of linear combinations of the columns of $\mathbf{X}$ orthogonal to $\mathbf{q}$ is equal to the set of linear combinations of the columns of $\mathbf{E}$. Deflation is the easiest way to add orthogonality constraints in many optimization problems encountered in multivariate analysis. In the usual consensus and hierarchical PCA methods, only a global component is introduced; the superblock is not explicitly introduced. Therefore deflation can only be considered on the blocks. Various deflation strategies (what to deflate and how) have been proposed:

a) Deflation on previous global components (Wold, Hellberg, Lundstedt, Sjöström, Wold, 1987). In this case, global components are uncorrelated, but block components do not belong to their block space and are correlated.

b) Deflation on previous block components (Wangen and Kowalski, 1989). In this case, block components belong to their block space and are uncorrelated, but global components are correlated.

c) Deflation on previous block loadings (Hassani, Hanafi, Qannari and Kohler, 2013). In this case, global components are uncorrelated; block components belong to their block space, but are correlated.

As a superblock is explicitly introduced in regularized consensus PCA, the deflation strategy is more flexible than in the consensus PCA and related methods. A fourth strategy can be considered: blocks and superblock are deflated on their own previous block and superblock components. Optimization problem (2) can be applied on the deflated blocks and the deflated superblock. However the deflated superblock is not equal to the concatenation of the deflated blocks and some properties of R-CPCA are lost when Mode B is used for the superblock. This strategy offers an attractive solution to the deflation problem encountered in consensus PCA: block components belong now to their block space and are uncorrelated and, in the same time, superblock components are uncorrelated.

# 5. Connection with regularized generalized canonical correlation analysis and PLS path modeling

Many multiblock component methods are special cases of RGCCA and PLS path modeling. In these methods, the superblock component $\mathbf{y}_{B+1}$ has various names as auxiliary variable, global component or superscore. These methods boil down to solve a stationary equation on $\mathbf{y}_{B+1}$ (see Table 3).

**Table 3:** Stationary equation for the global component for various multiblock component methods

| Special case of | Method | Reference | Stationary equation for the global component |
|---|---|---|---|
| RGCCA and R-CPCA | 1 | Covariance criterion (unweighted case) (Carroll, 1968b); PLS-PM on CPCA model with Mode A for all blocks and superblock and path weighting scheme (Lohmöller, 1989); CPCA (Westerhuis, Kourti, and MacGregor, 1998) | $\mathbf{y}_{B+1} \propto \mathbf{X}_{B+1} \mathbf{X}_{B+1}^t \mathbf{y}_{B+1}$ |
| | 2 | Generalized canonical correlation analysis (Carroll, 1968a); PLS-PM on CPCA model with Mode B for all blocks and superblock and factorial scheme (Lohmöller, 1989) | $\mathbf{y}_{B+1} \propto \sum_{b=1}^{B} \mathbf{X}_b \left( \mathbf{X}_b^t \mathbf{X}_b \right)^{-1} \mathbf{X}_b^t \mathbf{y}_{B+1}$ |
| | 3 | "Mixed" correlation and covariance criterion (unweighted case), (Carroll, 1968b) | $\mathbf{y}_{B+1} \propto \left( \sum_{b=1}^{B_1} \mathbf{X}_b \left( \frac{1}{n} \mathbf{X}_b^t \mathbf{X}_b \right)^{-1} \mathbf{X}_b^t + \sum_{b=B_1+1}^{B} \mathbf{X}_b \mathbf{X}_b^t \right) \mathbf{y}_{B+1}$ |
| R-CPCA | 4 | Hierarchical PCA (Smilde, Westerhuis, de Jong, 2003) | $\mathbf{y}_{B+1} \propto \sum_{b=1}^{B} \left\| \mathbf{X}_b^t \mathbf{y}_{B+1} \right\|^2 \mathbf{X}_b \mathbf{X}_b^t \mathbf{y}_{B+1}$ |
| PLS-PM | 5 | PLS-PM on CPCA model with Mode A for all blocks, Mode B for the superblock and centroid scheme (Wold, 1982) | $\mathbf{y}_{B+1} \propto \sum_{b=1}^{B} \left\| \mathbf{X}_b \mathbf{X}_b^t \mathbf{y}_{B+1} \right\|^{-1} \mathbf{X}_b \mathbf{X}_b^t \mathbf{y}_{B+1}$ |
| | 6 | PLS-PM on CPCA model with Mode A for all blocks, Mode B for the superblock and factorial scheme (Lohmöller, 1989); Hierarchical PCA-W (Smilde, Westerhuis, de Jong, 2003) | $\mathbf{y}_{B+1} \propto \sum_{b=1}^{B} \frac{\left\| \mathbf{X}_b^t \mathbf{y}_{B+1} \right\|^2}{\left\| \mathbf{X}_b \mathbf{X}_b^t \mathbf{y}_{B+1} \right\|^2} \mathbf{X}_b \mathbf{X}_b^t \mathbf{y}_{B+1}$ |
| | 7 | PLS-PM on CPCA model with Mode A for blocks and superblock and centroid scheme (Wold, 1982) | $\mathbf{y}_{B+1} \propto \mathbf{X}_{B+1} \mathbf{X}_{B+1}^t \sum_{b=1}^{B} \left\| \mathbf{X}_b \mathbf{X}_b^t \mathbf{y}_{B+1} \right\|^{-1} \mathbf{X}_b \mathbf{X}_b^t \mathbf{y}_{B+1}$ |
| | 8 | PLS-PM on CPCA model with Mode A for blocks and superblock and factorial scheme (Lohmöller, 1989) | $\mathbf{y}_{B+1} \propto \mathbf{X}_{B+1} \mathbf{X}_{B+1}^t \sum_{b=1}^{B} \frac{\left\| \mathbf{X}_b^t \mathbf{y}_{B+1} \right\|^2}{\left\| \mathbf{X}_b \mathbf{X}_b^t \mathbf{y}_{B+1} \right\|^2} \mathbf{X}_b \mathbf{X}_b^t \mathbf{y}_{B+1}$ |
| | 9 | Consensus PCA (Wold, Hellberg, Lundstedt Sjöström, Wold, 1987) | $\mathbf{y}_{B+1} \propto \sum_{b=1}^{B} \left\| \mathbf{X}_b^t \mathbf{y}_{B+1} \right\|^{-2} \mathbf{X}_b \mathbf{X}_b^t \mathbf{y}_{B+1}$ |

Note that the stationary equations for the global component in Methods 1 to 4 are special cases of equation (25). For Method 1, $m = 2$ and all $\mathbf{M}_b = \mathbf{I}$; for Method 2, $m = 2$ and all $\mathbf{M}_b = \frac{1}{n}\mathbf{X}_b^t\mathbf{X}_b$; for Method 3, $m = 2$, $\mathbf{M}_b = \frac{1}{n}\mathbf{X}_b^t\mathbf{X}_b$, $b = 1,...,B_1$ and $\mathbf{M}_b = \mathbf{I}$, $b = B_1+1,...,B$; for Method 4, $m = 4$ and all $\mathbf{M}_b = \mathbf{I}$. Therefore, Methods 1 to 4 are special cases of R-CPCA and consequently correspond to an optimization problem (Methods 1 to 3 are also special case of RGCCA which contains R-CPCA for $m = 1$ or 2). To our knowledge, the other methods in Table 3 do not correspond to any known optimization problems. Methods 5 to 8 are special case of PLS path modeling. It is worth noting that Hierarchical PCA-W is a special case of PLS-PM used on the CPCA model. When the configuration "Mode A for all blocks, Mode B for the superblock and factorial scheme" is selected, both methods yield the same block and superblock components. Note that RGCCA and PLS path modeling are available in the module PLS-PM of the XLSTAT software (Addinsoft, 2015). Method 9 corresponds to the first proposed consensus PCA method, but has some convergence problem. At that time, Methods 5 and 7 were available and were probably a better option as they have no convergence problem. In fact, it is really surprising that the PLS approach of Herman Wold has been so ignored in the papers related to consensus and hierarchical PCA. One objective of this paper was to give his deserved place to PLS-PM in multiblock component methods.

## 6. Conclusion

Many methods exist in the field of multiblock component methods. They are based on the construction of block components $\mathbf{y}_b = \mathbf{X}_b\mathbf{w}_b$. In this paper, we have mainly considered methods optimizing some criterion. Some methods are based on a correlation criterion and generalize canonical correlation analysis (Hotelling, 1936). Other methods are based on a covariance criterion and are extending inter-battery factor analysis (Tucker, 1958). Methods generalizing redundancy analysis (Van den Wollenberg, 1977) to more than two blocks have also been proposed. This distinction between these three sources leads to the following classification of multiblock component methods.

1)   *Correlation-based methods*

Main contributions for generalized canonical correlation analysis are found in Horst (1961a,b, 1965), Carroll (1968a), and Kettenring (1971).

2)   *Covariance-based methods*

Main contributions for extending Tucker's method to more than two blocks come from Carroll (1968b), Van de Geer (1984), Ten Berge (1988), Escofier and Pagès (1994), Chessel and Hanafi (1996), and Hanafi and Kiers (2006).

3)   *Mixed correlation- and covariance-based methods*

Carroll (1968b) proposed the "mixed" correlation and covariance criterion. Generalizations of redundancy analysis to the multiblock situation can be found in Van de Geer (1984) and in Tenenhaus and Tenenhaus (2011).

4) *Methods based on PLS path modeling*

Wold (1982, 1985) and Lohmöller (1989) applied the PLS path modeling (PLS-PM) approach to the "consensus PCA model". Hanafi (2007) showed the connections between this approach and generalized canonical correlation analysis.

5) *Methods based on PLS regression* (*consensus and hierarchical PCA*)

The consensus PCA algorithm proposed by Westerhuis, Kourti, and MacGregor (1998) is based on an iterative use of one-component PLS regressions. The other consensus and hierarchical PCA algorithms are quite similar; they differ on using other normalizations than the original one used in PLS regression. They are discussed in Wold, Hellberg, Lundstedt, Sjöström, and Wold (1987), Westerhuis, Kourti, and MacGregor (1998) and Smilde, Westerhuis, and de Jong (2003). Several consensus and hierarchical PCA methods are special case of PLS-PM applied to the "consensus PCA model" (see Table 3).

6) *Methods based on regularized generalized canonical correlation analysis* (*RGCCA*)

Using RGCCA (Tenenhaus and Tenenhaus, 2011, Tenenhaus and Guillemot, 2013) in the context of multiblock data has been discussed in details in Tenenhaus and Tenenhaus (2014). Many methods mentioned above are special cases of RGCCA.

7) *Methods for transforming several data matrices to maximal agreement*

The idea behind this approach is to find several components $\mathbf{y}_b^1 = \mathbf{X}_b\mathbf{w}_b^1, ..., \mathbf{y}_b^r = \mathbf{X}_b\mathbf{w}_b^r$ for each block in one step. Setting $\mathbf{W}_b = \left[\mathbf{w}_b^1, ..., \mathbf{w}_b^r\right]$, the transformed matrices are $\mathbf{X}_b\mathbf{W}_b$. Furthermore, the weight matrices $\mathbf{W}_b$ are constrained to be columnwise orthonormal. Van de Geer (1984), Ten Berge (1988) and Hanafi and Kiers (2006) proposed several correlation-based and covariance-based criteria for measuring the agreement between the transformed matrices $\mathbf{X}_b\mathbf{W}_b$. A very general monotonically convergent algorithm for finding the solutions for these methods has been proposed by Hanafi and Kiers (2006).

Regularized consensus PCA is a framework for many multiblock component methods. Many methods in Clusters 1, 2, 3 and 6 are recovered and those in Clusters 4 and 5 are much improved as regards their mathematical properties. Regularized consensus PCA is based on a well-defined optimization problem yielding a recurrence equation on the superblock component. The proposed iterative algorithm provides a fixed point of the recurrence equation. Furthermore, regularized consensus PCA allows more deflation strategies than in usual consensus PCA methods because a superblock is introduced in place of an auxiliary variable.

Several recent advances on RGCCA can be transposed to regularized consensus PCA. This concerns RGCCA for multigroup data (Tenenhaus and Tenenhaus, 2014), a sparse version of RGCCA (Tenenhaus, Philippe, Guillemot, Lê Cao, Grill, and Frouin, 2014), and a "kernel" extension of RGCCA (Tenenhaus, Philippe and Frouin, 2015).

# APPENDIX 1

## An algorithm for maximizing convex functions

We consider an objective function $\Psi(\mathbf{v}): \mathbb{R}^J \to \mathbb{R}$, convex and continuously differentiable and the following optimization problem:

(A1) $\qquad$ Maximize $\Psi(\mathbf{v})$ s.t. $\|\mathbf{v}\| = 1$.

We denote by $\nabla(\mathbf{v})$ the gradient of the objective function $\Psi(\mathbf{v})$. We suppose that $\|\nabla(\mathbf{v})\| > 0$ in the rest of this appendix. This assumption is not too binding as $\nabla(\mathbf{v}) = \mathbf{0}$ characterizes the *global minimum* of $\Psi(\mathbf{v})$. We consider the Lagrangian function $F(\mathbf{v}, \lambda) = \Psi(\mathbf{v}) - \frac{\lambda}{2}(\mathbf{v}^t \mathbf{v} - 1)$ related to optimization problem (A1). The optimal solution satisfies the stationary equation

(A2) $\qquad \dfrac{\partial F}{\partial \mathbf{v}}(\mathbf{v}) = \nabla(\mathbf{v}) - \lambda \mathbf{v} = \mathbf{0}$.

We denote a unit norm solution of the stationary equation (A2) by

(A3) $\qquad \mathbf{v}^* = \dfrac{\nabla(\mathbf{v}^*)}{\|\nabla(\mathbf{v}^*)\|}$.

Journée, Nesterov, Richtárik and Sepulchre (2010) proposed the monotone convergent algorithm for optimization problem (A1) given hereafter.

---

**Input:** $\mathbf{v}^0$ such that $\nabla(\mathbf{v}^0) \neq \mathbf{0}$, $\varepsilon$.
**Output:** $\mathbf{v}^s$ (approximate solution of (A1)).
**Repeat** ($s = 0, 1, 2, \ldots$):
(A4) $\qquad \mathbf{v}^{s+1} = f(\mathbf{v}^s) = \dfrac{\nabla(\mathbf{v}^s)}{\|\nabla(\mathbf{v}^s)\|}$
**Until** $\Psi(\mathbf{v}^{s+1}) - \Psi(\mathbf{v}^s) \leq \varepsilon$.

---

**ALGORITHM 2:** Gradient algorithm for problem (A1)

Note that any fixed point of the recurrence function $f$ is solution of the stationary equation (A2). We summarize the properties of Algorithm 2 in Proposition 2 below.

## Proposition 2

Let $\{\mathbf{v}^s\}$ be a sequence of unit norm vectors generated by the recurrence equation (A4). We assume that the norms of the gradients $\nabla(\mathbf{v}^s)$ are bounded from below: $\delta = \inf_s \|\nabla(\mathbf{v}^s)\| > 0$. The following properties hold:

a) The bounded sequence $\Psi(\mathbf{v}^s)$ is monotonically increasing and therefore converging.

b) All accumulation points of the sequence $\{\mathbf{v}^s\}$ are fixed points of $f$.

c) $\Psi(\mathbf{v}^s) \to \Psi(\mathbf{v}^*)$, where $\mathbf{v}^*$ is a fixed point of $f$.

d) $\|\mathbf{v}^{s+1} - \mathbf{v}^s\|^2 \leq \dfrac{2\left(\Psi(\mathbf{v}^{s+1}) - \Psi(\mathbf{v}^s)\right)}{\delta}$.

e) The sequence $\{\mathbf{v}^s\}$ is asymptotically regular: $\lim_{s \to 0} \|\mathbf{v}^{s+1} - \mathbf{v}^s\| = 0$.

f) Either $\{\mathbf{v}^s\}$ converges or the accumulation points of $\{\mathbf{v}^s\}$ form a continuum.

Before proving Proposition 2, we need to discuss various points and show some results.

## Discussion

Property (d) given in the proposition comes from Journée, Nesterov, Richtárik and Sepulchre (2010). The other properties can very simply be obtained by using the Fessler's version of Meyer's monotone convergence theorem (Meyer (1976), Fessler (2004)). An easy way to show that Algorithm 2 fulfills the conditions of Meyer's theorem is to explicit the MM approach (see Lange, 2013) used by Journée et al. Here MM stands for *maximization by minorization*. We use the property that a convex differentiable function lies above its linear approximation for any $\mathbf{v}$ and $\tilde{\mathbf{v}}$:

(A5) $\qquad \Psi(\tilde{\mathbf{v}}) \geq \Psi(\mathbf{v}) + \nabla(\mathbf{v})^t (\tilde{\mathbf{v}} - \mathbf{v})$.

A *minorizing* function $G(\tilde{\mathbf{v}}, \mathbf{v})$ is defined by considering the right term of (A.5):

(A6) $\qquad G(\tilde{\mathbf{v}}, \mathbf{v}) = \Psi(\mathbf{v}) + \nabla(\mathbf{v})^t (\tilde{\mathbf{v}} - \mathbf{v})$.

On the other hand, *maximizing* $G(\tilde{\mathbf{v}}, \mathbf{v})$ over $\tilde{\mathbf{v}}$ subject to $\|\tilde{\mathbf{v}}\| = 1$ is attained by choosing

(A7) $\qquad f(\mathbf{v}) = \dfrac{\nabla(\mathbf{v})}{\|\nabla(\mathbf{v})\|}$.

*Properties of* $G(\tilde{\mathbf{v}}, \mathbf{v})$

    a) $\Psi(\tilde{\mathbf{v}}) \geq G(\tilde{\mathbf{v}}, \mathbf{v})$.

    b) $G(\mathbf{v}, \mathbf{v}) = \Psi(\mathbf{v})$.

    c) Sandwich inequality:

(A8) $\quad\quad\quad\quad \Psi(\mathbf{v}) = G(\mathbf{v}, \mathbf{v}) \leq G(f(\mathbf{v}), \mathbf{v}) \leq \Psi(f(\mathbf{v}))$.

We now consider the sequence $\{\mathbf{v}^s\}$ generated by the recurrence equation (A4). We suppose that this sequence involves infinitely many distinct terms; otherwise the last given point satisfies $\mathbf{v}^s = f(\mathbf{v}^s)$ and therefore is a fixed point of $f$.

### Result 1

The bounded sequence $\Psi(\mathbf{v}^s)$ is monotonically increasing and therefore converging.

*Proof*

As the continuous function $\Psi$ is bounded on the unit sphere, this result is a direct consequence of the Sandwich inequality (A8).

### Result 2

Define a level set $C = \{\mathbf{v} : \|\mathbf{v}\| = 1 \text{ and } \Psi(\mathbf{v}) \geq \Psi(\mathbf{v}^0)\}$. The function $f : C \to C$ is uniformly compact on $C$.

*Proof*

We remind that a function $f$ is uniformly compact on $C$ if and only if there exists a compact set $K$ such that $f(\mathbf{v}) \in K$ for all $\mathbf{v} \in C$. As the level set $C = \{\mathbf{v} : \|\mathbf{v}\| = 1 \text{ and } \Psi(\mathbf{v}) \geq \Psi(\mathbf{v}^0)\}$ is compact, we can choose $K = C$. The Sandwich inequality (A8) implies $\Psi(\mathbf{v}^0) \leq \Psi(\mathbf{v}) \leq \Psi(f(\mathbf{v}))$ for all $\mathbf{v} \in C$ and therefore $f(\mathbf{v}) \in C$ for all $\mathbf{v} \in C$.

### Result 3

The function $f : C \to C$ is strictly monotone (increasing) with respect to $\Psi$. This means that the function $f$ satisfies two conditions:

- $\Psi(\mathbf{v}) \leq \Psi(f(\mathbf{v}))$, for any vector $\mathbf{v} \in C$.

- $\Psi(\mathbf{v}) < \Psi(f(\mathbf{v}))$ whenever $\mathbf{v}$ is *not* a fixed point of $f$, i.e. $f(\mathbf{v}) \neq \mathbf{v}$.

*Proof*

The first condition is deduced from the Sandwich inequality (A8). The second condition is shown by a contrapositive proof. If $\Psi(\mathbf{v}) = \Psi(f(\mathbf{v}))$, using (A8), we have $G(f(\mathbf{v}), \mathbf{v}) = G(\mathbf{v}, \mathbf{v})$. As $f(\mathbf{v})$ is the unique maximizer of $G(\tilde{\mathbf{v}}, \mathbf{v})$ over the set of unit norm vectors $\tilde{\mathbf{v}}$, we conclude that $\mathbf{v} = f(\mathbf{v})$ and therefore is a fixed point of $f$.

## Result 4

We assume $\delta = \inf_s \|\nabla(\mathbf{v}^s)\| > 0$. Then the following inequality holds:

$$(A9) \qquad \|\mathbf{v}^{s+1} - \mathbf{v}^s\|^2 \leq \frac{2\left(\Psi(\mathbf{v}^{s+1}) - \Psi(\mathbf{v}^s)\right)}{\delta}.$$

*Proof*

We deduce from (A8) and from the Proposition 3 of Journée et al. the following inequalities

$$(A10) \qquad \Psi(\mathbf{v}^{s+1}) - \Psi(\mathbf{v}^s) \geq G(\mathbf{v}^{s+1}, \mathbf{v}^s) - G(\mathbf{v}^s, \mathbf{v}^s) = \nabla(\mathbf{v}^s)^t (\mathbf{v}^{s+1} - \mathbf{v}^s) \geq \frac{\delta}{2} \|\mathbf{v}^{s+1} - \mathbf{v}^s\|^2$$

and therefore (A9).

**Proof of Proposition 2**

Point (a) corresponds to result 1. Points (b), (c), (e) and (f) come from a direct application of the Fessler's version of the Meyer's monotone convergence theorem. This theorem supposes three conditions which are satisfied here: (i) the function $f$ is continuous on $C$, (ii) the function $f$ is uniformly compact on $C$ and (iii) the function $f$ is strictly monotone (increasing) on $C$ with respect to the continuous function $\Psi$. Point (d) corresponds to result 4.

**Discussion**

At convergence of Algorithm A1, $\Psi(\mathbf{v}^{s+1}) - \Psi(\mathbf{v}^s) \leq \varepsilon$. This implies that $\|\mathbf{v}^{s+1} - \mathbf{v}^s\|^2 \leq 2\varepsilon/\delta$. When this condition is fulfilled, we consider that the obtained solution $\mathbf{v}^s$ is a fixed point of the recurrence equation (A4).